\documentclass{article}
\usepackage{amsmath,amsfonts,amssymb,amsthm,booktabs,color,epsfig,graphicx,url}
\usepackage[bookmarkstype=toc, colorlinks=true, linkcolor=cyan, pdfborder={0 0 0}, bookmarks=true, bookmarksnumbered=true]{hyperref}

\theoremstyle{definition}
\newtheorem{theorem}{Theorem}

\newtheorem{lemma}{Lemma}
\newtheorem{corollary}{Corollary}
\newtheorem{definition}{Definition}

\usepackage{algorithmic}
\usepackage{algorithm}

\usepackage{graphicx} 
\graphicspath{ {results} }
\usepackage[labelfont=bf]{caption}

\usepackage{authblk}

\title{Frank copula is minimum information copula under fixed Kendall's $\tau$}
\author[1]{Issey Sukeda}
\author[2]{Tomonari Sei}
\affil[1,2]{Graduate School of Information Science and Technology, The University of Tokyo}
\date{}

\begin{document}
\maketitle

\begin{abstract}
In dependence modeling, various copulas have been utilized. Among them, the Frank copula has been one of the most typical choices due to its simplicity. In this work, we demonstrate that the Frank copula is the minimum information copula under fixed Kendall's $\tau$ (MICK), both theoretically and numerically.
First, we explain that both MICK and the Frank density follow the hyperbolic Liouville equation. Moreover, we show that the copula density satisfying the Liouville equation is uniquely the Frank copula.
Our result asserts that selecting the Frank copula as an appropriate copula model is equivalent to using Kendall's $\tau$ as the sole available information about the true distribution, based on the entropy maximization principle.
\end{abstract}

\section{Introduction \label{sec:introduction}}

Copulas have gained popularity for modeling dependence in various fields. Numerous types and families of copulas have been proposed. Copulas without tail dependence, such as the Gaussian copula corresponding to the bivariate Gaussian distribution and the Frank copula, are typical examples. On the other hand, copulas exhibiting tail dependence, such as the Clayton copula, Gumbel-Hougaard copula, and Joe copula, are frequently preferred for datasets characterized by discernible tail dependence.

Among them, Frank copula has been used widely in applications due to its simple form including neuroscience~\cite{onken2009frank}, finance~\cite{rodriguez2007measuring}, and hydrology~\cite{favre2004multivariate}. Frank copula was originally proposed by Frank~\cite{frank1979simultaneous} as the solution of a functional equation regarding associativity, i.e., a copula $C(u,v)$ such that $C(u,v)$ and its survival copula $-1+u+v+C(1-u,1-v)$ are both associative simultaneously. In addition, Frank copula belongs to the Archimedean family, which is a very convenient class of copulas completely defined by a single function named \textit{generator}. 

\begin{definition}[Frank copula]
The cumulative distribution function of Frank copula is defined as
$$C^{Frank}_\theta(u,v) = \psi_\theta(\psi^{-1}_\theta(u) + \psi^{-1}_\theta(v)) = -\frac{1}{\theta} \ln \left(1+\frac{(e^{-\theta u}-1)(e^{-\theta v}-1)}{e^{-\theta}-1}\right), \ \theta \neq 0,$$
where $\psi$ denotes its \textit{generator function}:
$$\psi^{-1}_\theta(t) = -\ln{\left( \frac{e^{-\theta t} - 1}{e^{-\theta }-1}\right)}, \ \psi_\theta(t) = -\frac{1}{\theta}\ln{\left(1-(1-e^{-\theta})e^{-t} \right)}.$$
\end{definition}
\noindent Another characteristic is that Frank copula is comprehensive; that is,  as $\theta \to -\infty$ and  $\theta \to \infty$, it corresponds to the Frechet-Hoeffding lower bound and upper bound, respectively. In terms of tail dependence, Frank copula is known to be independent on both its lower tail near $(u,v) = (0,0)$ and its upper tail near $(u,v) = (1,1)$, i.e., $\lim_{u \to 0} \frac{C_\theta^{Frank}(u,u)}{u} = 0$ and $\lim_{u \to 1} \frac{1-2u+C_\theta^{Frank}(u,u)}{1-u} = 0$. 
While these characteristics are intuitive, the statistical interpretation of the Frank copula remains insufficient.

On the other hand, another class of copulas named the minimum information copula have been studied under several settings. 
This copula is defined as the optimal solution of the entropy maximization problem under preassigned constraints. As for the constraints, the linear constraint fixing the moments of the copula density into a certain constant has been the typical choice~\cite{bedford2014construction, butucea2015maximum, chen2024proper}, which includes fixing the Spearman's $\rho$ to a constant between 0 to 1~\cite{MEEU1997, pougaza2011maximum}. Different from these studies, Sukeda and Sei~\cite{sukeda2023minimum} considered the Kendall's $\tau$, a quadratic constraint with respect to the density function, as the given constraint. They provided several mathematical properties of this new class of copula, however its density function was not known explicitly. 
The relationship between the minimum information copula (or entropy maximization in other words) and the Frank copula has not been mentioned until this work.

Our work is largely devoted to prove that the Frank copula is indeed the minimum information copula under fixed Kendall's $\tau$\ (MICK) proposed by Sukeda and Sei~\cite{sukeda2023minimum} from both theoretical and numerical aspects. Our result provides a certain interpretation that choosing the Frank copula is equivalent to utilizing Kendall's $\tau$ as the only available information based on entropy maximization principle~\cite{jaynes1957information}. 
Our main theoretical result, stating that MICK and the Frank copula are identical, is presented in Section 2, followed by numerical justifications in Section 3. Finally, we discuss our interpretation on the use of Frank copula in Section 4.

\section{Frank copula and MICK coincides}
In this section, we state that MICK is identical to the Frank copula, whose density function is given as 
\begin{equation}
c^{Frank}_\theta(u,v) = \frac{\partial^2}{\partial u \partial v}C^{Frank}_\theta(u,v) = \frac{\theta(1-e^{-\theta})e^{-\theta(u+v)}}{[1-e^{-\theta}-(1-e^{-\theta u})(1-e^{-\theta v})]^2}, ~\label{eq:frank-density}
\end{equation}
which is parameterized by $\theta\ (\neq 0)$. 
The overall argument proceeds as follows: 
First, we confirm that both MICK and the Frank copula follow the same partial differential equation stating that the local dependence function~\cite{holland1987} is proportional to the original density function:
\begin{equation}\label{eq:pde-propto}
\frac{\partial}{\partial x}\frac{\partial}{\partial y}\log{p(x,y)} \propto p(x,y).
\end{equation}
Next, we highlight that this equation is the well-known Liouville equation in physics. 
Finally, by imposing the copula condition on the general solution of the Liouville equation, we demonstrate that the unique solution is the Frank copula density. The discussion overall supports the conclusion that MICK and Frank should represent the same distribution.

We start by showing that MICK satisfies \eqref{eq:pde-propto}. 
Let us define MICK as the optimal solution of the following information minimization (or entropy maximization) problem  with a constraint on Kendall's $\tau$:
\begin{equation}
\mathrm{minimize}\ \int_0^1 \int_0^1 p(u,v)\log{p(u,v)} \mathrm{d}u\mathrm{d}v, \label{prob:mick}
\end{equation}
$$\mathrm{s.t.}\ \int_0^1 p(u,v) \mathrm{d}u = 1,\ \int_0^1 p(u,v) \mathrm{d}v = 1,$$
$$0 < p(u,v),\ p\in C^2([0,1]^2),$$
$$\int_0^1 \int_0^1 \int_0^1 \int_0^1 \mathrm{sgn}(u-\tilde{u})\mathrm{sgn}(v-\tilde{v})p(u,v)p(\tilde{u},\tilde{v}) \mathrm{d}u\mathrm{d}v\mathrm{d}\tilde{u}\mathrm{d}\tilde{v} = \tau.$$
Note that Kendall's $\tau$ is represented via the density function. This problem is intractable due to its non-convexity and the uniqueness of the optimal solution is not trivial.
Unfortunately, even through the use of Lagrangian method, the explicit form of MICK cannot be obtained directly. However, the stationary condition that MICK should satisfy is tractable.
\begin{lemma} \label{lemma:mick-follows-pde}
MICK $p(x,y)$, if it exists, satisfies the following partial differential equation:
$$\frac{\partial}{\partial x}\frac{\partial}{\partial y}\log{p(x,y)} = 8\lambda p(x,y)$$
where $\lambda$ is the Lagrangian multiplier of the optimization problem~\eqref{prob:mick}.
\end{lemma}
\begin{proof}

Lagrangian function of the optimization problem is written as 
\begin{align*}
L[p] = \int_0^1 \int_0^1 p(x,y)\log{p(x,y)} \mathrm{d}x\mathrm{d}y &- \lambda (\int_0^1 \int_0^1 \int_0^1 \int_0^1 \mathrm{d}x\mathrm{d}y\mathrm{d}\tilde{x}\mathrm{d}\tilde{y}\ \mathrm{sgn}(x-\tilde{x})\mathrm{sgn}(y-\tilde{y})p(x,y)p(\tilde{x},\tilde{y})-\tau)\\
&-\int_0^1 dx \alpha(x)(\int_0^1 p(x,y) \mathrm{d}y - 1)-\int_0^1 dy \beta(y)(\int_0^1 p(x,y) \mathrm{d}x - 1),
\end{align*}
where $\lambda, \alpha, \beta$ are Lagrangian multipliers. 
The minimizer of the Lagrangian function can be found by variational calculus. 
Consider a infinitesimal change $\delta p(x,y)$. 
The first order variation of $L$ is 
\begin{align*}
L[p+\delta p(x,y)] - L[p] &= 
\int_0^1\int_0^1 \mathrm{d}x\mathrm{d}y \delta p(x,y) (\log{p(x,y)}+1) \\
&- \int_0^1 \int_0^1 \mathrm{d}x\mathrm{d}y \delta p(x,y)  \alpha(x)  - \int_0^1 \int_0^1 \mathrm{d}x\mathrm{d}y \delta p(x,y) \beta(y)\\
&- 2\lambda  \int_0^1\int_0^1 \mathrm{d}x\mathrm{d}y \int_0^1 \int_0^1 \mathrm{d}\tilde{x}\mathrm{d}\tilde{y} \delta p(x,y) \mathrm{sgn}(x-\tilde{x})\mathrm{sgn}(y-\tilde{y})p(\tilde{x},\tilde{y}) 
\end{align*}
Therefore, by taking a look inside $\int_0^1\int_0^1\mathrm{d}x\mathrm{d}y$,
\begin{align*}
    \log{p(x,y)}+1 &- \alpha(x)-\beta(y)\\
    &-2\lambda\left(\int_0^x \int_0^y \mathrm{d}\tilde{x}\mathrm{d}\tilde{y}p(\tilde{x},\tilde{y}) + \int_x^1 \int_y^1 \mathrm{d}\tilde{x}\mathrm{d}\tilde{y}p(\tilde{x},\tilde{y})-\int_x^1 \int_0^y \mathrm{d}\tilde{x}\mathrm{d}\tilde{y}p(\tilde{x},\tilde{y})-\int_0^x \int_y^1 \mathrm{d}\tilde{x}\mathrm{d}\tilde{y}p(\tilde{x},\tilde{y})\right)\\
    &= 0.
\end{align*}

\noindent Although it seems difficult to obtain the solution explicitly from it, by taking derivative of both sides, we obtain
$$\frac{\partial}{\partial x}\frac{\partial}{\partial y}\log{p(x,y)} = 8\lambda p(x,y)$$
from fundamental theorem of calculus.
\end{proof}

On the other hand, Frank copula is known to satisfy the same partial differential equation. The stronger result is obtained by Kurowicka and van Horssen~\cite{KUROWICKA2015127}, stating that the only Archimedean copula that satisfies this equation is the Frank copula. We do not use this result here.

\begin{lemma} \label{lemma:frank-follows-pde}
    Frank density with parameter $\theta$ satisfies the following equation:
    $$\frac{\partial^2}{\partial u \partial v}\log{c^{Frank}_\theta(u,v)} = 2\theta c^{Frank}_\theta(u,v).$$
\end{lemma}
\begin{proof}
The fact that the twice derivatives of log Frank density is proportional to the original Frank density can be confirmed by direct calculations as follows. First, we calculate the derivatives of log Frank density. Let us recall Frank density:
$$c_\theta^{Frank}(u, v) = \frac{\theta (1 - e^{-\theta})e^{-\theta(u + v)}}{\{1 - e^{-\theta} - (1 - e^{-\theta u})(1 - e^{-\theta v})\}^2}.$$
\begin{align*}
    \frac{\partial^2}{\partial u \partial v} \log{c_\theta^\mathrm{Frank}(u,v)} &= \frac{\partial^2}{\partial u \partial v} \log{e^{-\theta(u + v)}} -2\frac{\partial^2}{\partial u \partial v} \log{\{1 - e^{-\theta} - (1 - e^{-\theta u})(1 - e^{-\theta v})\}}\\
    &= -2\frac{\partial}{\partial u} \frac{- (1 - e^{-\theta u})(\theta e^{-\theta v})}{\{1 - e^{-\theta} - (1 - e^{-\theta u})(1 - e^{-\theta v})\}}\\
    &= 2\theta e^{-\theta v} \frac{(\theta e^{-\theta u})\{1 - e^{-\theta} - (1 - e^{-\theta u})(1 - e^{-\theta v})\} - (1 - e^{-\theta u})(- (1 - e^{-\theta v})(\theta e^{-\theta u}))}{\{1 - e^{-\theta} - (1 - e^{-\theta u})(1 - e^{-\theta v})\}^2}\\
    &= 2\theta^2 e^{-\theta u}e^{-\theta v}\frac{(1-e^{-\theta})}{\{1 - e^{-\theta} - (1 - e^{-\theta u})(1 - e^{-\theta v})\}^2}\\
    &= 2\theta \frac{\theta (1 - e^{-\theta})e^{-\theta(u + v)}}{\{1 - e^{-\theta} - (1 - e^{-\theta u})(1 - e^{-\theta v})\}^2}\\
    &= 2\theta c_\theta^{Frank}(u, v)
\end{align*}
\end{proof}

Next, we discuss the uniqueness. Since we observe that both MICK and Frank copula are solutions of the identical partial differential equation~\eqref{eq:pde-propto}, it suffices to guarantee the uniqueness of the solution. Let $\lambda\ (\in \mathbb{R})$ denote the constant of proportionality. Here, we point out that the log density in \eqref{eq:pde-propto} follows the hyperbolic Liouville equation~\cite{liouville1853}:
$$\frac{\partial^2}{\partial u \partial v}f(u,v) = \lambda \exp{(f(u,v))},$$
which is an important equation in the field of mathematics and physics with numerous generalization studies. 
The solution of this equation varies, however, we show that imposing the copula conditions on the solution leads to the unique solution, which coincides with the Frank density.

The most general solution to the hyperbolic Liouville equation is presented by Crowdy~\cite{crowdy1997general} as follows:
$$f(u,v) = \log{\frac{Y(u)W(v)}{\{c_1 Y_1(u)W_1(v) + c_2 Y_1(u)W_2(v) + c_3 Y_2(u)W_1(v) + c_4Y_2(u)W_2(v)\}^2}}$$
where $c_1c_4-c_2c_3 = -\frac{\lambda}{2}$ (we denote it as $-\theta$), $Y(u)$ and $W(v)$ are the Wronskians, i.e., $Y(u) = \frac{dY_1(u)}{du} Y_2(u) - \frac{dY_2(u)}{du} Y_1(u), W(v) = \frac{dW_1(v)}{dv} W_2(v) - \frac{dW_2(v)}{dv} W_1(v)$, and $Y_1, Y_2, W_1, W_2\in C^1$ are arbitrary. Denote
\begin{equation} \label{eq:def-of-F(u,v)}
F(u,v) = c_1 Y_1(u)W_1(v) + c_2 Y_1(u)W_2(v) + c_3 Y_2(u)W_1(v) + c_4Y(u)W_2(v)
= (Y_1(u),Y_2(u))\begin{pmatrix}
    c_1& c_2\\
    c_3& c_4
    \end{pmatrix}
    \begin{pmatrix}
    W_1(v)\\
    W_2(v)
    \end{pmatrix}
\end{equation}
for convenience. 
Note that the Frank density \eqref{eq:frank-density} is written in this form.

\begin{lemma} \label{lemma:uniqueness}
    Assume the logarithm of a bivariate copula density $p(u,v)$ satisfies the Liouville equation. Then, $p(u,v)$ is the Frank density.
\end{lemma}
\begin{proof}
    Assume $p(u,v) = \frac{Y(u)W(v)}{F(u,v)^2}$ is a copula density. 
    Since $\frac{1}{\lambda}\frac{\partial^2}{\partial u \partial v}\log{p(u,v)} = -\frac{1}{\theta}\frac{\partial^2}{\partial u \partial v}\log{F(u,v)} = -\frac{1}{\theta}\frac{-\frac{\partial}{\partial u}F(u,v)\frac{\partial}{\partial v}F(u,v)}{F(u,v)^2} = \frac{Y(u)W(v)}{F(u,v)^2}$, we have
    $$p(u,v) = -\frac{1}{\theta}\frac{\partial^2}{\partial u \partial v}\log{F(u,v)}.$$
    By integrating the both sides, the cumulative distribution function (cdf) can be calculated as 
    $$P(u,v) = -\frac{1}{\theta}\log{F(u,v)} + \alpha(u) + \beta(v).$$
    Here we can assume $P(u,v) = -\frac{1}{\theta}\log{F(u,v)}$ in general since $Y_1, Y_2, W_1, W_2$ can be arbitrary functions. This is because $P(u,v)$ stays the same form by replacing $Y_k(u)e^{-\theta \alpha(u)}$ with $Y_k(u)$ and $W_k(v)e^{-\theta \beta(v)}$ with $W_k(v)$ for $k=1,2$, respectively.
    The condition that $P(u,v)$ is a copula \footnote{Note that we do not use the 2-increasing condition.} can be formulated as 
    \[
     \begin{pmatrix}
      F(0,v)\\
      F(1,v)
     \end{pmatrix}
     = \underbrace{\begin{pmatrix}
      Y_1(0)& Y_2(0)\\
      Y_1(1)& Y_2(1)
     \end{pmatrix}\begin{pmatrix}
    c_1& c_2\\
    c_3& c_4
    \end{pmatrix}}_A
    \begin{pmatrix}
    W_1(v)\\
    W_2(v)
    \end{pmatrix}
    = \begin{pmatrix}
     1\\
     e^{-\theta v}
    \end{pmatrix}.
    \]
    Since the matrix $A$ is regular \footnote{By concatenating the equations for $v=0,1$, we obtain $A\begin{pmatrix}
        W_1(0)&W_1(1)\\
        W_2(0)&W_2(1)
    \end{pmatrix}=\begin{pmatrix}
        1&1\\
        1&e^{-\theta}
    \end{pmatrix}$. Since the right hand is obviously regular, $A$ is also regular.
    }
    , 
    \[
     \begin{pmatrix}
      W_1(v)\\
      W_2(v)
     \end{pmatrix}
     = A^{-1}\begin{pmatrix}
     1\\
     e^{-\theta v}
     \end{pmatrix}
    \]
    Similarly, for $u$ we have
    \[
     \begin{pmatrix}
      Y_1(u)\\
      Y_2(u)
     \end{pmatrix}
     = B^{-1}\begin{pmatrix}
     1\\
     e^{-\theta u}
     \end{pmatrix}
    \]
    hence \eqref{eq:def-of-F(u,v)} becomes
    \[
     F(u,v) = (1,e^{-\theta u})
     \underbrace{(B^{-1})^\top
     \begin{pmatrix}
     c_1& c_2\\
     c_3& c_4
     \end{pmatrix}
     A^{-1}}_{K}
     \begin{pmatrix}
     1\\
     e^{-\theta v}
     \end{pmatrix}.
    \]
    By taking $(u,v)=(0,0),(0,1),(1,0),(1,1)$, we obtain the conditions
    \[
     F(0,0) = (1,1)K\begin{pmatrix}1\\1\end{pmatrix} = 1,
     \quad F(0,1) = (1,1)K\begin{pmatrix}1\\e^{-\theta}\end{pmatrix} = 1,
    \]
    \[
     F(1,0) = (1,e^{-\theta})K\begin{pmatrix}1\\1\end{pmatrix} = 1,
     \quad F(1,1) = (1,e^{-\theta})K\begin{pmatrix}1\\e^{-\theta}\end{pmatrix} = e^{-\theta},
    \]
    which is summarized as 
    \[
     \begin{pmatrix}
      1& 1\\
      1& e^{-\theta}
     \end{pmatrix}
     K
     \begin{pmatrix}
     1& 1\\
     1& e^{-\theta}
     \end{pmatrix}
     =  \begin{pmatrix}
     1& 1\\
     1& e^{-\theta}
     \end{pmatrix}.
    \]
    Therefore, 
    \[
     K = \begin{pmatrix}
     1& 1\\
     1& e^{-\theta}
     \end{pmatrix}^{-1}
     = \frac{1}{e^{-\theta}-1}\begin{pmatrix}
     e^{-\theta}& -1\\
     -1& 1
     \end{pmatrix}.
    \]
    Once we obtain $K$, by \eqref{eq:def-of-F(u,v)}, $F$ is calculated as 
    \[
     F(u,v) = (1,e^{-\theta u})K\begin{pmatrix}1\\ e^{-\theta v}\end{pmatrix}
     = \frac{e^{-\theta}-e^{-\theta v}-e^{-\theta u}+e^{-\theta (u+v)}}{e^{-\theta} - 1}.
    \]
    Finally, the copula function is recovered as 
    $$P(u,v) = -\frac{1}{\theta}\log{F(u,v)} = -\frac{1}{\theta}\log{\left(1+\frac{(e^{-\theta u}-1)(e^{-\theta v}-1)}{e^{-\theta}-1}\right)},$$
    which is obviously the Frank copula.

\end{proof}

From Lemma~\ref{lemma:mick-follows-pde}, Lemma~\ref{lemma:frank-follows-pde}, and Lemma~\ref{lemma:uniqueness} combined altogether, we conclude that MICK and Frank copulas are identical to each other as our main result.

\begin{theorem}\label{thm:param}
MICK and Frank copula with parameter $\theta$ are identical under the relationship $8\lambda = 2\theta$ where $\lambda$ is the Lagrangian multiplier in Lemma~\ref{lemma:mick-follows-pde}.
\end{theorem}


Moreover, it is known that the relationship between Kendall's $\tau$ and the parameter $\theta$ of the Frank copula is given by using Debye function~\cite{mcneil2015quantitative}:
$$\tau = 1-\frac{4}{\theta}[1-D_1(\theta)]$$
$$D_k(x) = \frac{k}{x^k}\int_0^x \frac{t^k}{e^t-1} dt.$$
\noindent Therefore, we can also view the relationship between the Kendall's $\tau$ of MICK and the Frank parameter $\theta$.

\begin{corollary}
MICK with parameter $\tau$ and Frank copula with parameter $\theta$ are identical under the following relationship:
    \begin{align}
      \tau = 1-\frac{4}{\theta}[1-\frac{1}{\theta}\int_0^\theta \frac{t}{e^t-1} dt], \label{eq:tauvstheta}  
    \end{align}
    where $\tau$ is a constant in the optimization problem of MICK and $\theta$ is the parameter of Frank copula.
\end{corollary}

\section{Numerical confirmation}
\begin{figure}[htbp]
\begin{minipage}[b]{0.5\linewidth}
    \centering
    \includegraphics[width=7cm]{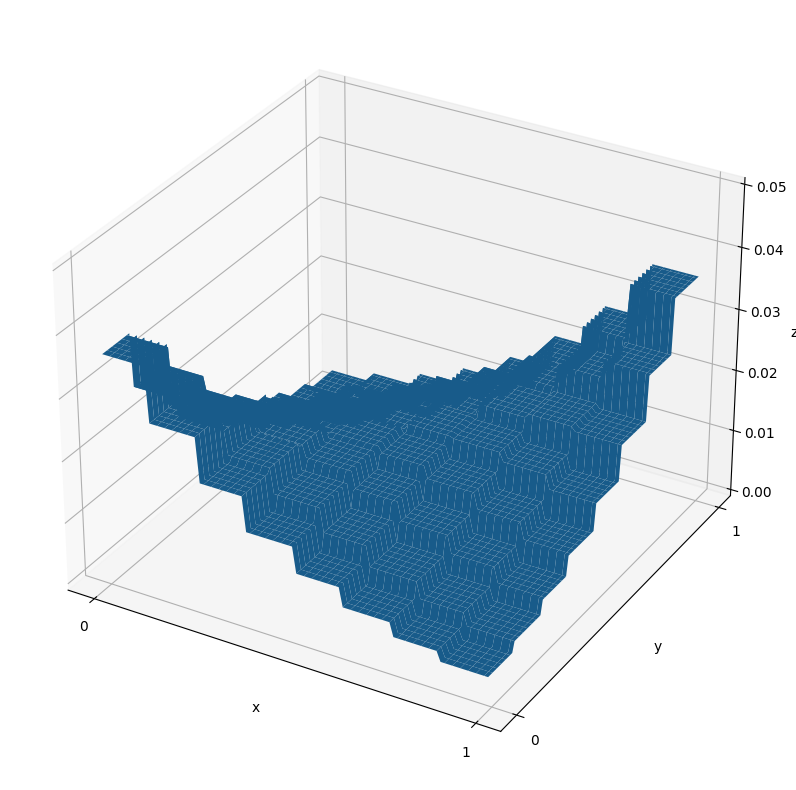}
  \end{minipage}
  \begin{minipage}[b]{0.5\linewidth}
    \centering
    \includegraphics[width=7cm]{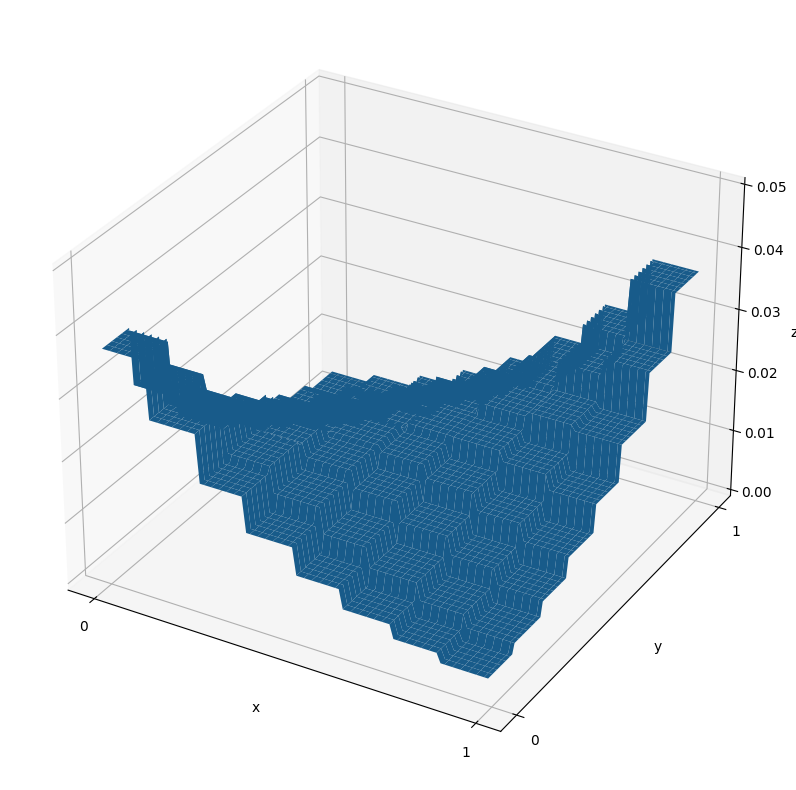}
  \end{minipage}
  \caption{left : checkerboard approximation of Frank density with $\theta=3$ / right : MICK ($\tau$=0.307) with gridsize $8\times8$ }
  \label{fig:checkerboardmick}
\end{figure}

The statement in Theorem~\ref{thm:param} is confirmed by numerical calculation presented in Figure~\ref{fig:checkerboardmick}. 
In numerical calculations, we could utilize the checkerboard approximation, which has uniform density within each square grids $[\frac{i-1}{n},\frac{i}n] \times [\frac{j-1}{n},\frac{j}n]\ (i, j = 1, \dots, n)$, where $n \times n$ denotes the gridsize. 

\begin{figure}[htbp]
    \centering
    \includegraphics[width=7cm]{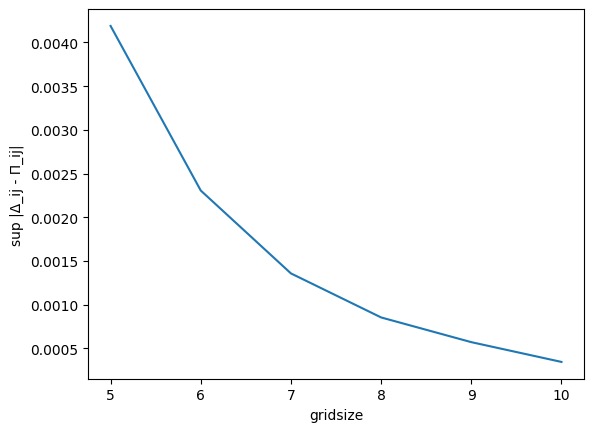}
  \caption{Illustration of the convergence for the checkerboard MICK}
  \label{fig:error_wrt_grid}
\end{figure}

First, we plot the checkerboard approximation~\cite{durrleman2000copulas} of Frank density with the parameter $\theta=3$ in the left side of Figure~\ref{fig:checkerboardmick}, which is formulated as
$$\Delta_{ij} = C_\theta^{\mathrm{Frank}}\left(\frac{i}{n},\frac{j}{n}\right) - C_\theta^{\mathrm{Frank}}\left(\frac{i-1}{n},\frac{j}{n}\right) - C_\theta^{\mathrm{Frank}}\left(\frac{i}{n},\frac{j-1}{n}\right) + C_\theta^{\mathrm{Frank}}\left(\frac{i-1}{n},\frac{j-1}{n}\right).$$
On the other hand, the right plot in Figure~\ref{fig:checkerboardmick} is the checkerboard MICK $\Pi_{ij}$, which is obtained by solving the discrete optimization problem using Scipy. See Sukeda and Sei~\cite{sukeda2023minimum} for details. Here we set $\tau=0.307$ as the constraint so that $\theta\ (=0.3)$ and $\tau$ satisfies the relationship \eqref{eq:tauvstheta}. For both plots, the gridsize is set to $8 \times 8$.
We observe that these two figures, the checkerboard MICK and the checkerboard Frank, look identical. Moreover, the same holds true when the parameters $\theta$ and $\tau$ is set otherwise as long as they satisfies \eqref{eq:tauvstheta}. 
Furthermore, the value $\sup_{i,j} |\Delta_{ij} - \Pi_{ij}|$ is plotted in Figure~\ref{fig:error_wrt_grid}, showing empirically that the checkerboard MICK approaches the checkerboard Frank copula as the gridsize grows. Also, note that the bivariate function defined by $\mathfrak{C}_n(u,v) = n^2\sum_{i=1}^n\sum_{j=1}^n \Delta_{ij} \int_0^u \chi_{i,n}(x) dx \int_0^v \chi_{j,n}(y) dy$ is known to converge to the Frank copula $C_\theta^{\mathrm{Frank}}$~\cite{durrleman2000copulas}.

\section{Conclusion}

While Frank copula is useful and can be characterized by its associativity and tail independence, its statistical interpretation has been insufficient. Our result, the Frank copula is the MICK, provides a new interpretation to the use of Frank copula. That is, opting for the Frank copula is equivalent to choosing the most entropic (or natural) copula while utilizing the value of the true Kendall's $\tau$ as an only available information in advance. 

However, the relationship between the associativity of Frank copula and our result remain unknown. Its difficulty comes from the fact that associativity is defined for the cumulative distribution function of the copula, while our result is based on the Shannon entropy, which is defined for copula densities. Filling this gap could be one of the future directions. Another direction of study could be replacing the Shannon entropy with Tsallis entropy or $q$-entropy, which should lead to different copulas.

\bibliographystyle{abbrv}
\bibliography{References_1-2}
\appendix

\end{document}